# Preventing Corrosion of Aluminum Metal with Nanometer-Thick Films of $Al_2O_3$ Capped with $TiO_2$ for Ultraviolet Plasmonics

Prithu Roy,[1] Clémence Badie,[2] Jean-Benoît Claude,[1] Aleksandr Barulin,[1] Antonin Moreau,[1] Julien Lumeau,[1] Marco Abbarchi,[3] Lionel Santinacci,[2] Jérôme Wenger[1,*]

[1] Aix Marseille Univ, CNRS, Centrale Marseille, Institut Fresnel, AMUTech, Marseille, France

[2] Aix Marseille Univ, CNRS, CINaM, AMUTech, Marseille, France

[3] Aix Marseille Univ, Université de Toulon, CNRS, IM2NP, AMUTech, Marseille, France

* Corresponding author: jerome.wenger@fresnel.fr

**Abstract**

Extending plasmonics into the ultraviolet range imposes the use of aluminum to achieve the best optical performance. However, water corrosion is a major limiting issue for UV aluminum plasmonics, as this phenomenon occurs significantly faster in presence of UV light, even at low laser powers of a few microwatts. Here we assess the performance of nanometer-thick layers of various metal oxides deposited by atomic layer deposition (ALD) and plasma-enhanced chemical vapor deposition (PECVD) on top of aluminum nanoapertures to protect the metal against UV photocorrosion. The combination of a 5 nm $Al_2O_3$ layer covered by a 5 nm $TiO_2$ capping provides the best resistance performance, while a single 10 nm layer of $SiO_2$ or $HfO_2$ is a good alternative. We also report the influence of the laser wavelength, the laser operation mode and the pH of the solution. Properly choosing these conditions significantly extends the range of optical powers for which the aluminum nanostructures can be used. As application, we demonstrate the label-free detection of streptavidin proteins with improved signal to noise ratio. Our approach is also beneficial to promote the long-term stability of the aluminum nanostructures. Finding the appropriate nanoscale protection against aluminum corrosion is the key to enable the development of UV plasmonic applications in chemistry and biology.

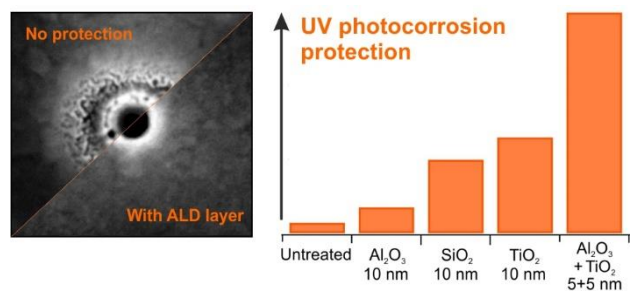

Figure for Table of Content

**Introduction**

Ultraviolet plasmonics is a promising new range of research to push nano-optics forward and take maximum advantage of the strong molecular absorption in the 200-300 nm range.[1–4] Applications involve surface-enhanced Raman spectroscopy,[5–10] protein autofluorescence detection,[11–16] and catalysis.[17–20] Working in the UV range imposes to rethink the choice of classical metals used for plasmonics, as gold and silver have interband transitions and strong losses below 400 nm.[21,22] So far, aluminum is best material choice for UV plasmonics, as it features relatively low losses, has a low intrinsic cost and is CMOS compatible.[23–25] Another interest of aluminum is that its plasmonic optical response covers a very broad spectral range from the deep UV up to the infrared, enabling multicolor applications in spectroscopy,[26,27] fluorescence sensing,[28,29] light harvesting [30,31] and optical color filters.[32,33]

While aluminum has many advantages, a major drawback concerns its limited stability in presence of water.[31,34–37] Bulk aluminum is pretty reactive and will dissolve quickly while reacting with water.[34] Although the alumina layer naturally present at the surface partially protects the metal against water corrosion, the oxide layer exhibits cracks at the junction between metal grains where pitting corrosion can occur.[38,39] Therefore, corrosion will limit the long-term durability of aluminum-based devices.[40] Owing to the large surface to volume ratio of plasmonic nanostructures, the corrosion phenomenon becomes even more restrictive at the nanoscale.[36,37] In the context of ultraviolet plasmonics, the situation is further degraded as UV light has been shown to further accelerate the corrosion process.[41–43] UV illumination has two main effects: (i) while increasing the growth of aluminum oxide, it also increases the porosity and the presence of cracks where pitting corrosion occurs,[43,44] and (ii) UV irradiation of water promotes the formation of radicals which dramatically accelerate the dissolution of alumina and aluminum.[42,45–47]

Various approaches have been explored to protect aluminum against corrosion,[34,35] for instance by passivating the aluminum surface with polyvinylphosphonic acid and polydopamine layers.[42,48,49] However, as we show in the Supporting Information Fig. S1 and S2, the use of organic molecules is limited in the context of UV plasmonics due to the direct photodamage on the organic polymers by the UV light and the photogeneration of reactive radicals. Atomic layer deposition (ALD) and plasma-enhanced chemical vapor deposition (PECVD) of thin inert layers of metal oxides offer powerful alternative approaches for corrosion protection owing to their ability to deposit dense layers of oxide materials with controlled nanometer thickness and conformal coating at the nanoscale.[50] Alumina ($Al_2O_3$),[51–53] titania ($TiO_2$) [54–56] and hafnia ($HfO_2$) [57,58] are commonly used to protect flat metal surfaces. However, the performance of these coatings has never been tested in the newly developing context of UV plasmonics, *i. e.* in the presence of UV irradiation and the photogenerated radical species, and while keeping the constraint of nanoscale features for the plasmonic nanodevices.

Here we assess the UV photocorrosion protection performance of nanometer-thick layers of various metal oxides deposited by ALD or PECVD on top of aluminum nanoapertures. Individual circular nanoapertures of 65 nm diameter milled in a 100 nm thick aluminum film form a simple and reproducible platform to check the nanostructure photostability in presence of water and UV irradiation.[42,59] Beyond the choice of the best material for the nanometer capping layer, we also discuss the influence of other experimental conditions like the laser wavelength, its operation mode or the pH of the solution. As soon as water is present, aluminum UV plasmonics will face the photocorrosion problem. Therefore finding the optimal protection method at the nanoscale is crucial for the future development of aluminum plasmonics and its applications in chemistry and molecular biology. As additional advantage of our approach, our results are easy to reproduce and do not involve complex or uncommon materials or procedures.

**Experimental Section**

*Nanoaperture fabrication.* A 100 nm-thick layer of aluminum is deposited on top of cleaned quartz coverslips using electron-beam evaporation (Bühler Syrus Pro 710). The deposition parameters (chamber pressure $10^{-6}$ mbar, deposition rate 10 nm/s) are set to minimize the amount of oxides into the bulk aluminum layer.[23,60] Nanoapertures with 65 nm diameter are then milled using gallium-based focused ion beam (FEI dual beam DB235 Strata, voltage 30 kV, ion current 10 pA).[61]

*ALD deposition.* The $Al_2O_3$, $HfO_2$, $TiO_2$ thin films were grown by ALD in a Fiji 200 reactor (Veeco/Cambridge Nanotech) operating with Ar as gas carrier. These materials are commonly used for ALD,[51–58] ensuring that our results can be easily reproduced by others. However, their application for aluminum plasmonics has never been reported so far. The reaction chamber was maintained at 150°C The deposition conditions have been set according to previous work.[62–64] $Al_2O_3$ was grown from Trimethylaluminium (TMA from Strem Chemicals, 98%) and deionized water ($\rho$ = 18.2 MΩ·cm) that were stored in dedicated canisters at room temperature (RT). The ALD cycle consisted of sequential pulse and purge of TMA and $H_2O$, alternatively. The pulse and purge durations were 0.06:10 s for both precursors. $HfO_2$ layers were deposited from Tetrakis(ethylmethylamino)hafnium (TEMAH from Strem chemicals, 99.999%) and $O_2$ (Linde Electronics, 99.996%) by plasma-enhanced ALD (PE-ALD). The TEMAH was stored at 115°C. The PE-ALD cycle consisted of sequential pulse and purge of TEMAH and $O_2$, alternatively. The pulse and purge durations were 1:5 s and 6:5 s for TEMAH and $O_2$, respectively. Plasma-assistance was used during the $O_2$ injection using a ICP remote plasma set at P = 300 W. $TiO_2$ films were grown from Tetrakis(dimethylamino)titanium (TDMAT, Strem chemicals, 99%) and deionized water by thermal ALD. The TDMAT was heated at 85°C and water was left at RT. The ALD cycle consisted of sequential pulse, exposure and purge of both TDMAT and $H_2O$, alternatively. The pulse, exposure and purge durations were 1:7:15 s for the Ti precursor and 0.2:7:15 s for water.

*PECVD deposition.* PECVD treatment was performed using a PlasmaPro NGP80 from Oxford Instruments. After a pumping step of 20 minutes at 300°C, the $SiO_2$ layer is deposited at the same temperature using two percursors: 5% $SiH_4/N_2$ and $N_2O$ with flows of 160 and 710 sccm respectively. The chamber pressure is 1000 mTorr and high-frequency power of 20W. The deposition rate of silica is 1 nm/s. For the UV/ozone post-PECVD treatment, we use a Novascan PSD-UV cleaner with a 100 W mercury lamp.

*Experimental setup.* Our home built confocal microscope features three different ultraviolet laser sources: a pulsed picosecond 266 nm laser (Picoquant LDH-P-FA-266, 70 ps pulse duration, 80 MHz repetition rate), a pulsed picosecond 295 nm laser (Picoquant VisUV-295-590, 70 ps pulse duration, 80 MHz repetition rate) and a CW 266 nm laser (CryLas FQCW266-10-C). The lasers are used separately from each other. The laser beams are expanded to reach a 5 mm diameter before the LOMO 58x 0.8 NA water immersion microscope objective. The spot size at the focus of the objective has a diameter of 500 nm. To monitor the occurrence of corrosion, we measure the increase in the transmission in the 310-410 nm spectral band through the nanoaperture using a microLED illumination (Zeiss 423053-9071-000). A 50 µm pinhole is conjugated to the microscope focus for spatial filtering and ensuring only the region of interest around the nanoaperture is monitored. A photomultiplier tube (Picoquant PMA 175) with a photon counting module (Picoquant Picoharp 300) records the transmitted intensity after spectral filtering to select the 310 – 410 nm range (Semrock FF01-300/LP-25 and FF01-375/110-25).

*FCS experiments on streptavidin.* Powdered Streptavidin (Streptomyces avidinii, M=52.8 kDa, 24 tryptophan residues) is dissolved in pH-7 Hepes buffer (100mM NaCl, 25mM Hepes, 0.5 v/v% Tween 20). The streptavidin concentration is set to 10 µM and controlled by a spectrophotometer (Tecan Spark 10M). To improve the UV photostability, our measurement buffer includes GODCAT oxygen scavenger (10 wt% D-glucose, 0.83 µM catalase and 0.3 µM glucose oxidase) and 10 mM DABCO (1,4-Diazabicyclo[2.2.2]octane) antifading agent as previously described.[16] Each FCS trace is recorded for 100 s integration time. Fitting of the correlation functions is performed with a standard Brownian diffusion model:[15]

$$G(\tau) = \frac{\langle F(t)F(t+\tau)\rangle}{\langle F(t)\rangle^2} - 1 = \frac{1}{N_{mol}} \left(1 - \frac{B}{F_{PMT}}\right)^2 \frac{1}{(1+\frac{\tau}{\tau_d})\cdot(1+\frac{\tau}{\kappa^2\tau_d})^{0.5}} \quad (1)$$

$B/F_{PMT}$ is the ratio of background counts B to the total detected fluorescence signal $F_{PMT}$, $N_{mol}$ is the number of detected molecules, $\tau_d$ is the diffusion, and κ is the aspect ratio of the detection volume which is kept at a constant value of 1 according to our earlier works[15,59,65]. At 5 µW excitation power, the background B amounts to 1.2 kcounts/s while it increases to 2.7 kcounts/s at 20 µW. the background stems mostly from the GODCAT fluorescence, and it saturates at high power.

**Results and Discussion**

We assess the corrosion resistance of a single nanoaperture of 65 nm diameter milled into a 100 nm thick aluminum layer, which serves as a generic platform for UV aluminum plasmonics.[15,59] The nanoaperture is filled with a water solution and is exposed to a focused UV laser spot of increasing power (Fig. 1a). To monitor the occurrence of photocorrosion, we simultaneously record the transmission through the nanoaperture. This transmission is highly sensitive to small changes in the aperture diameter, allowing to monitor the photocorrosion evolution *in situ* and in real time.

In the absence of any extra protective layer, we find that the nanoaperture quickly corrodes even at a low UV power of 10 µW within a few tens of seconds (Fig. 1b,c). The aluminum surrounding the aperture is largely dissolved, leading to the emergence of a porous region at the boundary of the untouched aluminum film. On the contrary, when a 5 nm-thick $Al_2O_3$ layer followed by a 5 nm-thick $TiO_2$ capping layer are deposited by ALD, the nanoaperture withstands a 10× higher laser fluence for an extended time period (Fig. 1b). This demonstrates the interest for appropriate photocorrosion protection by covering the aluminum nanostructure with a nanometer-thick films of metal oxides. It also shows that direct laser-induced photodamage is not present in our observations (with our pulsed laser at 266 nm, we estimate that the photodamage threshold occurs at 150 µW average power).[66]

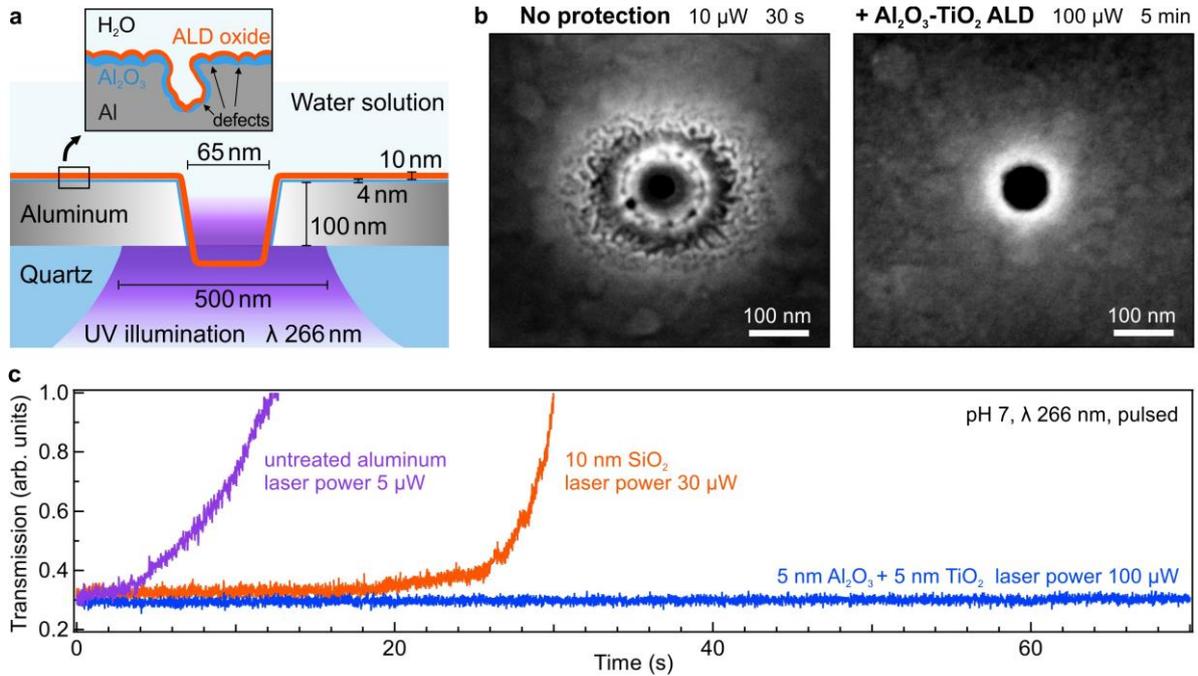

**Figure 1.** (a) Single aperture of 65 nm diameter milled in a 100 nm thick aluminum film. The aperture is illuminated by a focused ultraviolet laser beam, and a water solution fills the aperture and the upper medium. The natural aluminum oxide (alumina) layer covers the aluminum bulk, but this layer is not fully dense or homogeneous on the surface, leaving defect sites for aluminum corrosion. The nanometer-thick extra oxide layer efficiently protects the aluminum surface. (b) Example of scanning electron microscopy images of a single aperture without and with a protective ALD layer (here 5 nm of $Al_2O_3$ followed by 5 nm of $TiO_2$ as capping). In the absence of protection, corrosion is clearly visible around the central aperture. (c) Transmitted intensity time trace for different nanoaperture samples and UV laser power. The stability of the time trace indicates corrosion resistance.

To assess the photocorrosion resistance of a sample, we introduce the damage threshold power $P_{th}$ as the maximum laser power for which the transmission signal is stable for at least 90 s. For optical powers higher than $P_{th}$, the aperture will corrode within less than 90 s, leading to a rapid increase in the transmission signal (as in Fig. 1c for instance). We use this definition of $P_{th}$ as a benchmark to assess the influence of the oxide material deposited by ALD and PECVD (Fig. 2a) and other experimental parameters (Fig. 2b-e). All the aluminum nanoapertures follow the same treatment before the protective layer deposition process. They all feature the same typical 3-4 nm natural aluminum oxide layer. To enable a straightforward comparison between the different materials, we use the same thickness of 10 nm for all our experiments. Larger thicknesses could further improve the corrosion resistance as it was previously reported in the context of flat metal surfaces.[52,55,57,58] However, thicknesses above 15 nm are not appropriate for plasmonic applications which require nanoscale features and short distances to the metal film so as to maximize the electromagnetic field enhancement.[1] On the lower side, we found that a 5 nm thickness did not provide a sufficient improvement in the photocorrosion protection. Therefore, we selected a constant 10 nm as a trade-off between corrosion protection and plasmonic performance. As compared to an unprotected aluminum nanoaperture,[15] the presence of the 10 nm protective layer did not significantly alter the plasmonic performance: the experimental fluorescence enhancement factor recorded on p-terphenyl molecules inside a 65 nm aluminum aperture was reduced by less than 10% in presence of a supplementary 10 nm-thick metal oxide layer.

While all the different extra layers show an improvement of the photocorrosion resistance as compared to the raw aluminum aperture (Fig. 2a), there are significant differences among the materials. $Al_2O_3$, $TiO_2$ and $HfO_2$ and their combinations were deposited by ALD while $SiO_2$ was deposited by PECVD (see Methods section for details). The 10 nm-thick $Al_2O_3$ layer offers the lowest improvement in the protection resistance among the materials tested. The cracks and residual porosity of this material together with its limited resistance to water dissolution explain this behaviour.[38,39,51] The other oxides $SiO_2$, $TiO_2$ and $HfO_2$ feature a better protection performance with a higher threshold power. The best overall protection is brought by the combination of 5 nm $Al_2O_3$ layer followed by a 5 nm $TiO_2$ capping layer. In this case, our experiments are limited by the direct laser damage of the aluminum structure at 150 µW.[66] Our observation of the best performance for the $Al_2O_3$ layer with a $TiO_2$ capping goes in line with the earlier works which found that these two layers provided the best barrier to protect copper against water corrosion.[51,55] Surprisingly, the combination of 5 nm $Al_2O_3$ layer followed by a 5 nm $HfO_2$ layer does not provide an improved corrosion resistance despite the good performance of the 10 nm $HfO_2$ layer alone. This might be due to the limited 5 nm thickness and the residual porosity of the $HfO_2$ layer which may not be enough to fully protect against pitting corrosion in the $Al_2O_3$ layer cracks.

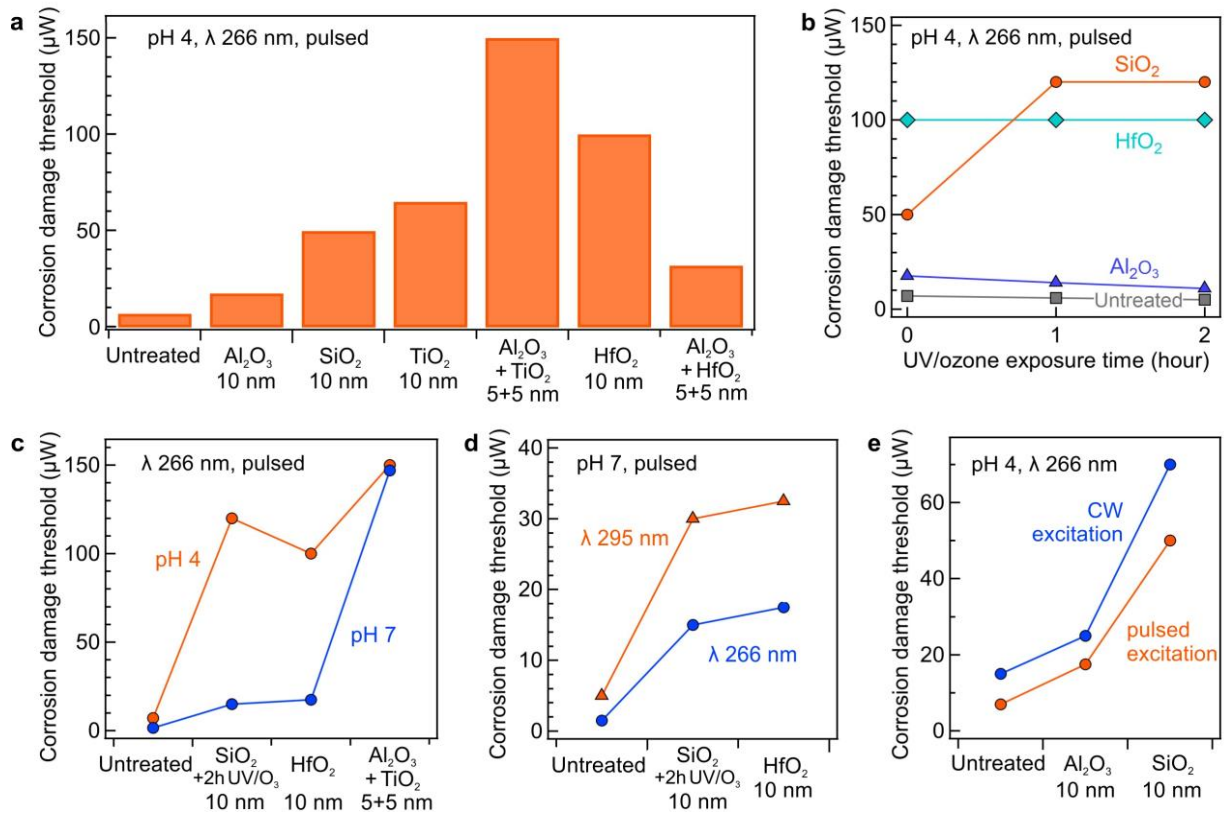

**Figure 2.** Corrosion protection strategies and influence of experimental parameters. Throughout this figure, the upper line indicates the pH, laser wavelength and laser operation mode (pulsed or continuous). (a) Threshold power leading to corrosion in 90 s as a function of different oxides deposited on top of the aluminum surface. $Al_2O_3$, $TiO_2$ and $HfO_2$ were deposited by ALD while $SiO_2$ was deposited by PECVD. (b) Influence of supplementary UV/ozone exposure on different samples. (c) Influence of the solution pH, a more acidic pH enables a better corrosion protection. The $Al_2O_3+TiO_2$ case is limited by the available laser power here and the occurrence of direct laser-induced damage. (d) Influence of the laser wavelength. (e) Influence of the illumination mode: pulsed or continuous (CW).

Beyond the material choice and its thickness, there are many other experimental parameters to explore. In Fig. 2b we assess the influence of an additional exposure of the oxide layer to UV mercury lamp in a UV/ozone cleaner. During the UV/ozone treatment with a mercury lamp (100W, emission wavelengths 185 nm and 254 nm), the oxygen present in the air reacts to form ozone, which in turn improves the corrosion and dissolution resistance of $SiO_2$.[67–69] One hour of UV/ozone treatment significantly improves the corrosion resistance of the $SiO_2$ layer from 50 to 120 μW. On the contrary, we found that while UV/ozone has no noticeable effect on the resistance performance of the hafnia and titania layers, it degrades the resistance of the alumina layer. For $Al_2O_3$, the UV/ozone exposure increases the density of available or unoccupied aluminum sites, which will increase the dissolution rate.[41,70]

The pH of the solution also plays a critical role in the photocorrosion (Fig. 2c), as a lower pH reduces the dissolution rate.[42,69] Switching from pH 7 to pH 4 significantly increases the corrosion threshold power for the different materials tested. For the combined $Al_2O_3$ + $TiO_2$ layer, the performance is limited by the direct laser photodamage of aluminum at 150 µW, but it is remarkable that this layer protects efficiently the aluminum even at pH 7 condition.

The generation rate of radicals by UV light in water depends on the illumination wavelength.[45] Changing the laser wavelength from 266 to 295 nm improves the corrosion resistance by above two-fold (Fig. 2d), which goes in line with the change in the water photolysis rate reported previously.[45] Lastly, we also tested the operation mode of the 266 nm laser, switching from the pulsed 70 ps 80 MHz repetition rate to a continuous wave (CW) mode. With the same average power, the picosecond pulses can induce multi-photon dissociation and ionization process in water,[42,45] while the CW laser only leads to one-photon absorption processes. Our observations show an improved corrosion resistance while using the CW laser as compared to the picosecond laser (Fig. 2e). However, the UV photocorrosion resistance is not infinitely high for the CW laser, indicating that one-photon absorption also participate to initiate the photocorrosion in addition to multiphoton processes.

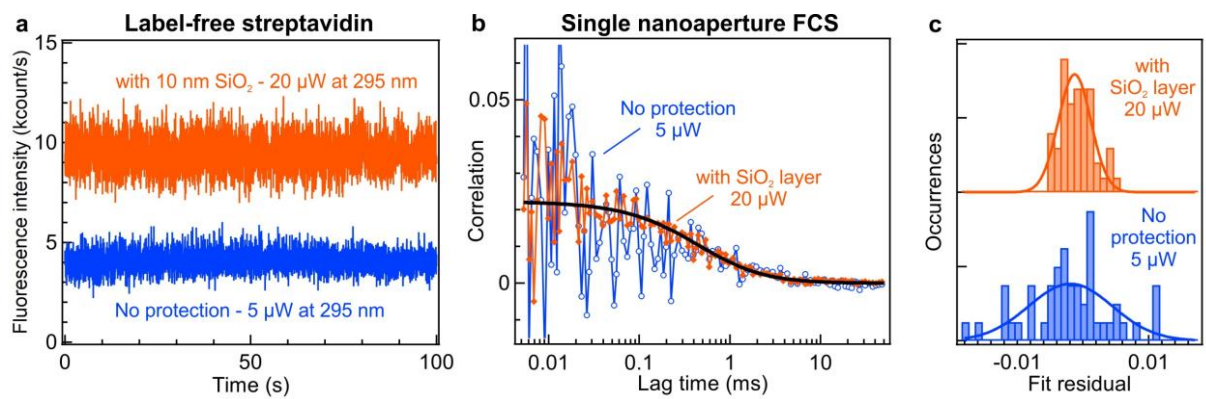

**Figure 3.** Application of the corrosion protection of single aluminum nanoapertures for the ultraviolet autofluorescence detection of label-free streptavidin proteins. (a) Ultraviolet (310-410 nm) autofluorescence time traces for a single 65 nm diameter aperture filled with a 10 µM solution of streptavidin at pH 7. The fluorescence stems from the 24 tryptophan aminoacid residues present in each streptavidin tetramer. (b) Temporal correlation of the time traces in (a), the thick black line is a numerical fit. Thanks to the higher 20 µW excitation power enabled by the extra $SiO_2$ protection layer, the noise is significantly reduced. Fitting the FCS data yields $N_{mol}$ = 26 ± 2 at 5 µW and $N_{mol}$ = 25 ± 1 at 20 µW. The diffusion times are respectively 630 ± 100 µs at 5 µW and 670 ± 50 µs at 20 µW. (c) Histogram of the fit residuals in (b) for the two excitation powers, the curves are vertically offset for clarity.

We use the photocorrosion strategy outlined in Fig. 2 to improve the detection of label-free streptavidin proteins (Fig. 3). Each streptavidin tetramer contains 24 natural tryptophan residues which are fluorescent in the ultraviolet range and can be detected using single aluminum nanoapertures combined with a UV confocal optical microscope.[15,16] However, in the absence of additional protective layer, the UV photocorrosion of aluminum limits the maximum usable power to a few microwatts only. With the 295 nm laser, the maximum power is 5 µW to avoid damaging the nanoaperture (Fig. 2d). In this condition, fluorescence correlation spectroscopy (FCS) can still be recorded, but the curve is very noisy and FCS parameters such as the number of proteins, their brightness or diffusion time cannot be reliably extracted (Fig. 3b,c). At higher powers with an untreated sample, the aperture quickly deteriorates as shown in Fig. 1c, making the experiment impossible. Thanks to a 10 nm $SiO_2$ layer with UV/ozone treatment, the corrosion resistance of the aluminum sample is significantly improved so that the laser power can be increased to 20 µW and no photocorrosion is observed (Fig. 3a). This higher excitation power yields a larger fluorescence brightness, which directly improves the quality of the FCS data and reduces the noise (Fig. 3b,c).[65] This set of data demonstrates that the application of the corrosion protection strategies derived from Fig. 2 enable the detection of label-free proteins using an aluminum plasmonic nanostructure in the UV. As aluminum is necessary for UV plasmonics and water is ubiquitous for biochemistry and molecular biology applications, the corrosion protection will be central to many other applications. Additionally, we show in the Supporting Information Fig. S3 that the photocorrosion protection developed here for UV plasmonics is also useful to protect the aluminum structures in presence of a corrosive buffer solution as often used for protein denaturation studies. The approaches used here can thus be extended to other uses of aluminum plasmonics, they are not restricted only to UV.

**Conclusions**

The enhanced photocorrosion rate of aluminum while exposed to UV light is a major bottleneck limiting the applications of aluminum UV plasmonics. Here, we have explored how nanometer-thick layers of metal oxides can help in protecting the aluminum surface against the UV photocorrosion. The combination of a 5 nm $Al_2O_3$ layer covered by a 5 nm $TiO_2$ capping provided the best resistance performance, both in acidic and neutral pH. Using a 10 nm $SiO_2$ capping with a 1 hour UV/ozone post-treatment is a good alternative. The ease of reproducibility and the use of common materials are additional advantages of our approach to further generalize the use of aluminum in plasmonics. In addition to finding the best nanomaterial for the protective capping, the photocorrosion resistance can be tuned by setting the experimental conditions to pH 4, using a CW laser and a higher illumination wavelength around 300 nm. Moreover, the 5 nm $Al_2O_3$ layer with 5 nm $TiO_2$ capping also provides an excellent protection against long-term exposure to corrosive chloride solutions. This enables protein denaturation studies and other biophysical works using aluminum nanostructures for an extended time period while reusing the same aluminum nanostructures. By bringing new data on the aluminum photocorrosion process as well as efficient nanomaterial solutions, this work makes a significant advance to enable the future development of UV plasmonic applications in chemistry and biology.

**Supporting Information**

The Supporting Information is available free of charge on the ACS Publications website at DOI:

Polyvinylphosphonic acid and polydopamine as photocorrosion protection, ageing of UV microscope objective, aluminum corrosion protection in presence of chloride buffer

**Funding sources**

This project has received funding from the European Research Council (ERC) under the European Union's Horizon 2020 research and innovation programme (grant agreement No 723241) and from the Excellence Initiative of Aix-Marseille University - A*MIDEX, a French "Investissements d'Avenir" program. We acknowledge the NanoTecMat platform of the IM2NP institute of Marseille.

**Conflict of Interest**

The authors declare no competing financial interest.

**Supporting Information for**

**Preventing Corrosion of Aluminum Metal with Nanometer-Thick Films of $Al_2O_3$ Capped with $TiO_2$ for Ultraviolet Plasmonics**

Prithu Roy,[1] Clémence Badie,[2] Jean-Benoît Claude,[1] Aleksandr Barulin,[1] Antonin Moreau,[1] Julien Lumeau,[1] Marco Abbarchi,[3] Lionel Santinacci,[2] Jérôme Wenger[1,*]

[1] Aix Marseille Univ, CNRS, Centrale Marseille, Institut Fresnel, AMUTech, Marseille, France

[2] Aix Marseille Univ, CNRS, CINaM, AMUTech, Marseille, France

[3] Aix Marseille Univ, Université de Toulon, CNRS, IM2NP, AMUTech, Marseille, France

* Corresponding author: jerome.wenger@fresnel.fr

**Contents:**

## S1. Polyvinylphosphonic acid and polydopamine as photocorrosion protection

In an earlier work, some of the present authors used a surface passivation based on the organic molecules polyvinylphosphonic acid (PVPA) and polydopamine (PDA) to protect against UV photocorrosion.[1] Figure S1 compares the results of this past approach with the current results achieved using a nanometer-thick conformal oxide layer deposited by ALD or PECVD.

The corrosion threshold powers previously published in [1] cannot be directly compared to the present data because of two major changes in our experimental setup. First, the transmission of the microscope objective has been modified significantly. Previously, we used a Zeiss Ultrafluar 0.6NA whose transmission got degraded over time (Supporting Information Fig. S2). We have solved this issue here using a LOMO 0.8NA objective featuring with 3x larger transmission. Second, in our previous work, we mainly investigated apertures of large 160 nm diameter featuring lower plasmonic enhancement and hence less photoinduced corrosion. The case of 65 nm diameter apertures is more challenging, but it also corresponds to the main application for UV plasmonics (Fig. 3).

To easily compare the protections brought by PVPA and PDA polymers with those achieved using nanometer-thick conformal oxide layers, we have performed a new set of experiments using the same LOMO objective and the same 65 nm diameter aperture (Fig. S1). The organic polymer layers have been deposited using the protocols described in [1] which build on earlier reports for PVPA [2] and PDA.[3,4] Using these protocols, the PVPA layer thickness is about 5 nm,[2] while the PDA layer is 15 nm thick.[3]

Figure S1 summarizes our results regarding the photocorrosion stability. Using the 266 nm pulsed laser and a pH 4 buffer, the PVPA-coated aperture withstands up to 18 µW laser power without showing clear signs of photocorrosion. For PDA, the corrosion threshold power moves to 30 µW. At pH 7, the corrosion resistance is reduced to 8 µW for PVPA and 9 µW for PDA. Comparing these levels with the conformal oxide layers deposited by ALD and PECVD, it appears that the photocorrosion protection offered by organic polymers is clearly overwhelmed. The new approach developed here using nanometer-thick films of $Al_2O_3$ capped with $TiO_2$ provides a better solution. Altogether, this supplementary data clarifies the positioning of our work respective to the state-of-the-art and clearly motivates the nanomaterials deposition by ALD and PECVD in the context of UV plasmonics.

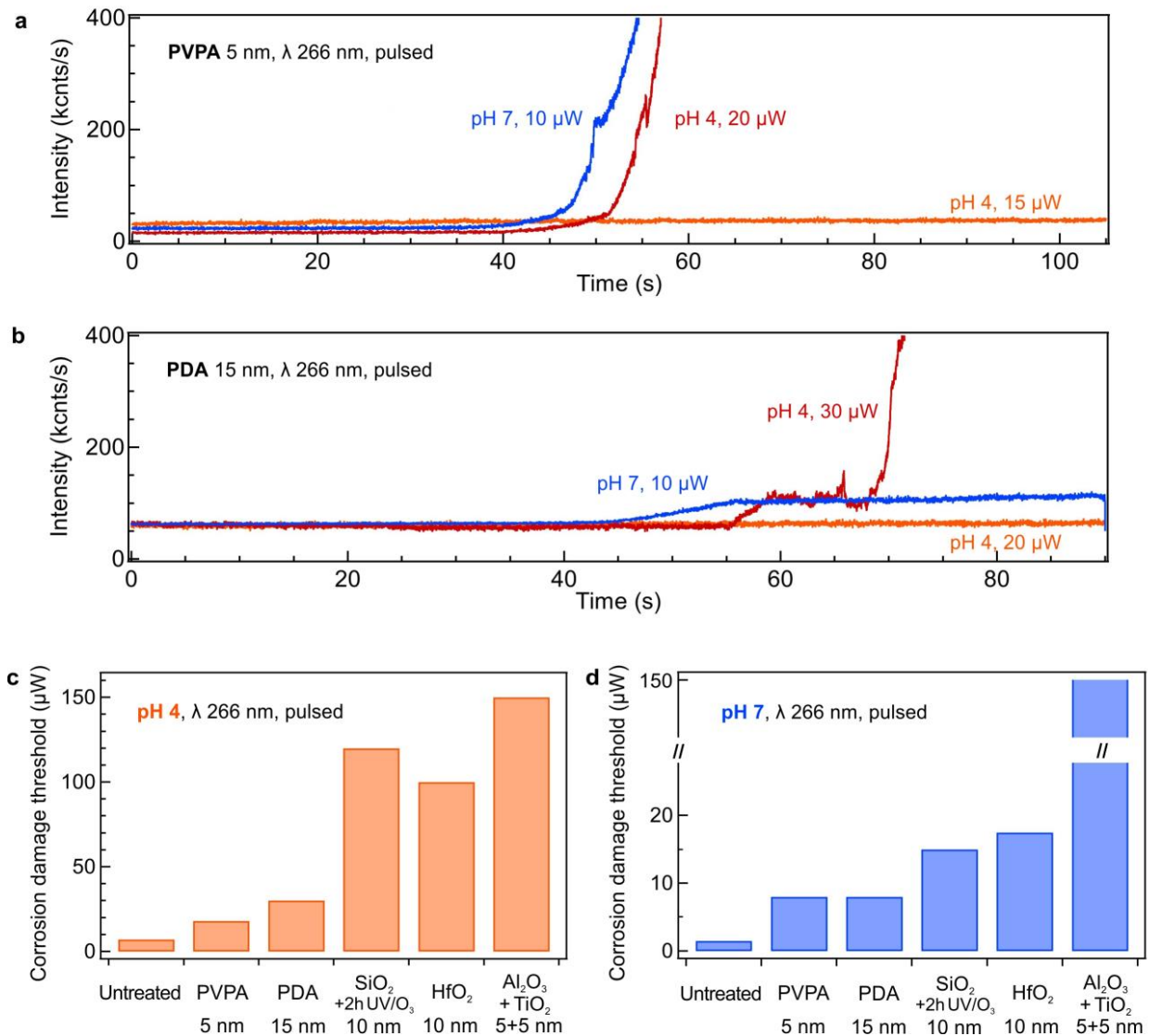

**Figure S1.** Comparison between metal oxide layers and organic polymers as UV photocorrosion for aluminum. (a,b) Transmitted intensity time trace for different 65 nm nanoaperture samples and UV laser power. The stability of the time trace indicates corrosion resistance while the sudden exponential growth in transmission is the sign of corrosion occurrence. In (a) the aluminum surface is covered by 5 nm PVPA and in (b) with 15 nm PDA. (c,d) Comparison of the threshold powers leading to corrosion in 90 s as a function of different layers deposited on top of the 65 nm aluminum nanoaperture for pH 4 (c) and pH 7 (d).

## S2. Ageing of UV microscope objective

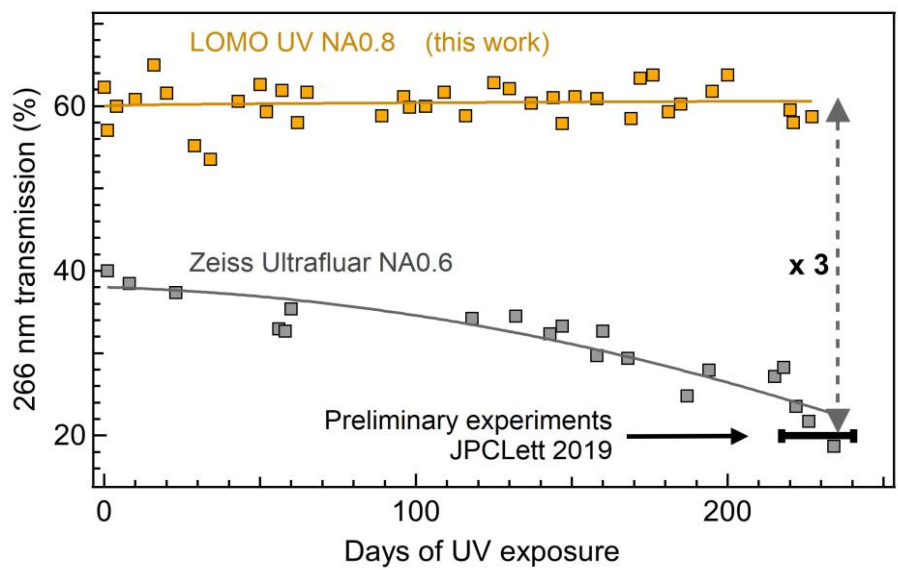

**Figure S2.** Comparison between the 266 nm transmission of the LOMO objective used in this work and the ZEISS ultrafluar objective used in our previous study.[1] While the ZEISS objective transmission tend to degrade over time, the LOMO objective appears to remain free of this effect. A possible explanation for this phenomenon is the degradation of silicone glues used in the ZEISS objective upon UV irradiation.[5] While comparing between the present experiments and some of our past results, the transmission of the microscope objective has increased by 3 times, leading to a higher UV intensity and thus a faster photocorrosion rate.

## S3. Aluminum corrosion protection in presence of chloride buffer

The protection strategies developed here using ALD for UV plasmonics can also turn useful to protect the aluminum structures in presence of a corrosive buffer solution. Biophysical studies of protein unfolding often use high molar concentrations of guanidinium chloride (GdmCl) to serve as a denaturant for proteins.[6,7] Because of the high chloride ions content of GdmCl, this compound is highly corrosive for aluminum. Here, we test our different ALD layers to record their ability to protect the aluminum layer against the long-term corrosion by GdmCl. The samples are immersed in a de-ionized water solution with 6 M of GdmCl and 500 mM of NaCl at pH 7, which is representative of the highest concentrations used to denaturate proteins and peptides.[6,7] Without any corrosion protection, the aluminum film starts to be corroded within a few hours and becomes significantly damaged after 5 days of immersion (Fig. S3). Using the combined 5 nm $Al_2O_3$ + 5nm $TiO_2$ or 5 nm $Al_2O_3$ + 5nm $HfO_2$ layers significantly improves the corrosion resistance up to 3 weeks immersion into the GdmCl solution. Remarkably, these combined layers feature much less defects than the layer containing only $SiO_2$. This set of results illustrates another interest of ALD to protect aluminum against corrosion and enable biophysical studies in harsh environments containing a high chloride concentration of several molar.

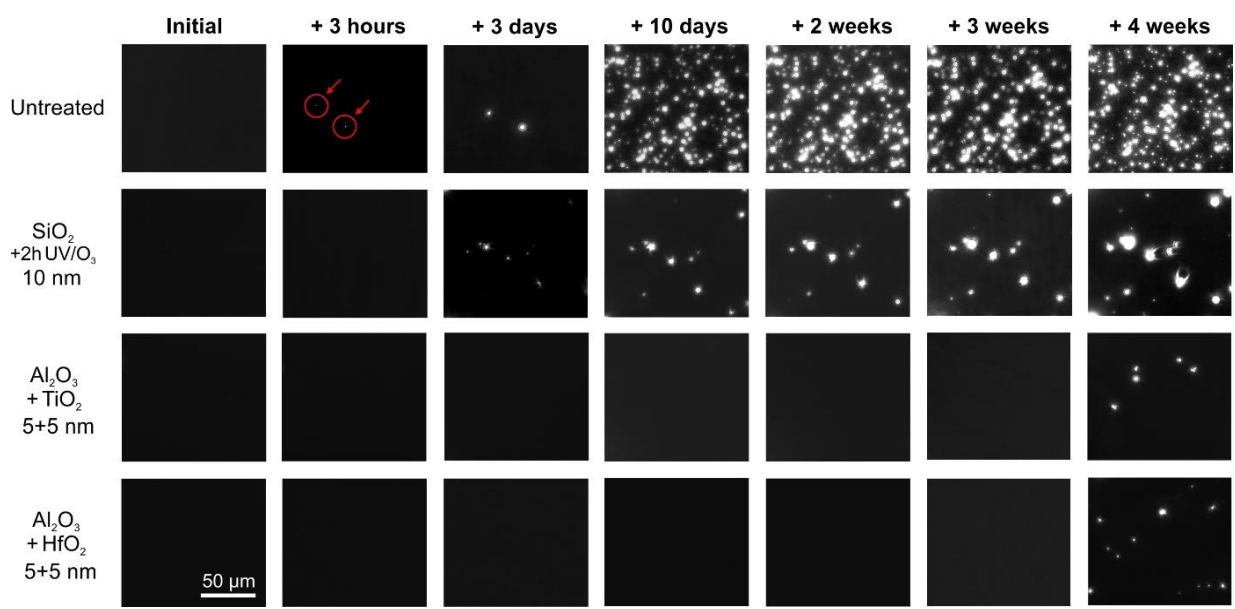

**Figure S3.** Transmission optical microscope images of the aluminum film during several days of immersion into a water solution containing 6 M of guanidinium chloride (GdmCl) and 500 mM of sodium chloride (NaCl) at pH 7. Fiducials located outside the camera field of view were used to image the same zone throughout the experiments.